\documentclass[prb,twocolumn,showpacs,amsmath,amssymb]{revtex4}

\usepackage{graphicx}
\usepackage{psfrag}  
\usepackage{dcolumn}
\usepackage{bm}
\usepackage{multirow} 
\usepackage{threeparttable} 
\usepackage{hyperref} 
\hypersetup{backref=true,
 pdfnewwindow=true, colorlinks=true,
 linkcolor=blue, anchorcolor=blue,
 citecolor=blue, filecolor=blue,
 menucolor=blue, urlcolor=blue}
\def\pitem[#1]{\textsuperscript{#1\hspace{0.1cm}}\ignorespaces}

\def\phAlpha{$P4_12_1 2$}    
\def\phAlphaSch{$D^4_4$}     
\def\phAlphaEnan{$P4_32_1 2$}
\def\phBeta{$I\bar{4}2d$}    
\def\phBetaSch{$D^{12}_{2d}$} 
\def\phIdeal{$Fd\bar{3}m$}   
\def\phIdealSch{$O^7_h$}     
\def\phSubGr{$P2_12_12_1$}   
\def\phSubGrSch{$D_2^4$}     
\def\phAlBTwo{$C222_1$}      
\def\phAlphaScott{\phAlBTwo} 
\def\phAlphaScottSch{$D_2^5$}

\def\ac{a_{\rm c}}

\newcolumntype{.}{D{.}{.}{-1}} 
\newcolumntype{,}{D{.}{ }{-1}} 
\newcommand{\mc}[1]{\multicolumn{1}{c@{}}{#1}}

\def\dvm#1{\marginpar{\small DV: #1}}
\def\scm#1{\marginpar{\small SC: #1}}
\def\dvm#1{}
\def\scm#1{}

\begin{document}

\title{Structural stability and lattice dynamics of SiO$_2$ cristobalite}

\author{Sinisa Coh} 
\email{sinisa@physics.rutgers.edu} 
\author{David Vanderbilt} 
\email{dhv@physics.rutgers.edu}
\affiliation{
  Department of Physics \& Astronomy, Rutgers University, Piscataway,
  NJ 08854-8019, USA
}

\date{\today}

\begin{abstract}  
Among the phases of SiO$_2$ are $\alpha$- and $\beta$-cristobalites,
which have a long and somewhat controversial history
of proposed structural assignments and phase-transition
mechanisms. Recently, Zhang and Scott found new indications that
the higher-temperature $\beta$ phase has space group \phBeta\ and,
by assuming a group-subgroup relationship between phases, they
argued that the lower-temperature $\alpha$ phase should have lower
symmetry than that of the widely-accepted \phAlpha\ space group.
With this motivation, we use first-principles calculations to
investigate the energy, structure, and local stability of
\phAlpha\ and \phBeta\ structures.  We also compute the frequencies
of the zone-center phonon modes in both structures, as well as certain 
zone-boundary modes in the \phBeta\ structure, and compare with
experiment.  We then argue that the various \phAlpha\ and
\phBeta\ enantiomorphs can be grouped into three clusters, each of
which is identified with a three-dimensional manifold of structures
of \phSubGr\ symmetry in which the \phAlpha\ and \phBeta\ appear as
higher-symmetry special cases.  We find that there are relatively
high energy barriers between manifolds, but low barriers within a
manifold.
Exploring the energy landscape within one of these manifolds,
we find a minimal-energy path connecting \phAlpha\ and \phBeta\
structures with a surprisingly low barrier of $\sim$5\,meV per
formula unit.  Possible implications for the phase-transition
mechanism are discussed.
\end{abstract}


\maketitle

\section{Introduction}
\label{sec:intro}

The fact that SiO$_2$ can exist in numerous crystalline and amorphous
forms, and its status as one of the most prevalent compounds on earth,
has stimulated a long history of experimental and theoretical
investigation.  Here we focus on the $\alpha$ (``low'') and $\beta$
(``high'') cristobalite phases, which are stable near the
melting temperature and are metastable at room temperature.

The structure of the higher-temperature $\beta$ phase has a history
of controversy.  Early indications of a cubic structure with
180$^\circ$ bond angles (space group \phIdeal)\cite{wyckoff} were
challenged by others \cite{peacor,leadbetter} who hypothesized
that the
true $\beta$-phase structure has lower symmetry and that the
apparent cubic structure arises from averaging over spatial domains
or dynamical fluctuations.  In particular, Wright and
Leadbetter\cite{leadbetter} argued for a tetragonal structure
belonging to space group \phBeta~(\phBetaSch).
While some subsequent works have provided
support for this identification,\cite{okeeffe,liu-prl93,scott}
other authors maintain that it is better to describe the $\beta$ phase as
a dynamically disordered one having overall \phIdeal\
symmetry but with a large population of rigid-unit-mode (RUM)
fluctuations.\cite{dove,dove-comm}  To some degree, the argument
may be semantic; if the fluctuations have strong short-range
correlations in space and time, it is difficult to distinguish this
picture from one of dynamic domains of a lower-symmetry structure.
\cite{liu-resp}  Thus, for example, either picture may be able to
explain the fact that there are two more first-order lines in the Raman
and infrared spectra than would be expected from
\phIdeal\ symmetry,\cite{scott} and the question of which description
is ``correct'' might depend on the time and length scales of the
experimental probes in question.

In contrast, the assignment of the $\alpha$-cristobalite phase to the
tetragonal \phAlpha\ ($D_4^4$) space group\cite{pluth} has until
recently been noncontroversial.  However, based
on a reexamination of Raman and infrared vibrational spectroscopies,
Zhang and Scott\cite{scott} have recently raised new questions about
the identity of the $\alpha$ phase.  By using Raman spectroscopy to
study small single crystals of $\beta$-cristobalite, these authors
argued that the $\beta$ structure must be $D_{2d}$, not cubic,
and assuming a group-subgroup relationship for the $\beta$-to-$\alpha$
transition, concluded that the $\alpha$ phase should have some lower
symmetry such as $D_2$ instead of $D_4$.  The
apparent $D_4$ symmetry of $\alpha$-cristobalite could result from
spatial or dynamic averaging over $D_2$ domains, in analogy to what
had been proposed for the $\beta$ phase.  To support their assumption
that a group-subgroup relationship should hold,
Zhang and Scott also pointed to the temperature dependence of
the optical phonon frequencies near the phase transition as being
inconsistent with a reconstructive phase transition\cite{toledano}
and as suggesting a nearly second-order behavior, although arguing in
the opposite direction are the facts that
the latent heat and volume change at the
transition are quite substantial.\cite{dove-private}

In their paper, Zhang and Scott\cite{scott}
reexamined earlier Raman and infrared spectroscopic measurements
not only on the $\alpha$- and $\beta$-cristobalite
SiO$_2$,\cite{swainson} but also on $\alpha$ and $\beta$
AlPO$_4$ (Ref.~\onlinecite{nicola})
and $\alpha$ BPO$_4$ (Ref.~\onlinecite{dultz}) cristobalites.
Note that the replacement of Si atoms by Al and P (or B and P)
atoms
immediately reduces the symmetry according to
\phAlpha~(\phAlphaSch) $\rightarrow$ \phAlBTwo~($D_2^5$)
for the $\alpha$ phase\cite{explan-symm}
and \phBeta~(\phBetaSch) $\rightarrow$ $I\bar{4}$~($S_4^2$) for the
$\beta$ phase.  Also of possible relevance is the pressure-induced
phase transition from $\alpha$-SiO$_2$ to a high-pressure monoclinic
cristobalite phase.\cite{monoclinic}
The relationship of these other cristobalites to the $\alpha$
and $\beta$ phases of SiO$_2$ is an interesting avenue for future
exploration, but falls outside the scope of the present work.

First-principles calculations of the structural and lattice dynamical
properties of SiO$_2$ have a long and productive history.
\cite{allan-87,allan-90,liu-prl93,liu-prb94,tse-prb95,demkov-prb95,
  hamann-prl96,demuth,catti-jpcb00,civalleri-ms02,donadio-prb03,
  chagarov-prb05,flocke-jpcb05,martonak-nm06,yu-prb07} While quite a
few of these works specifically address the ${\alpha}$-cristobalite
structure, \cite{allan-87,allan-90,catti-jpcb00,donadio-prb03,
  demuth,liu-prb94,yu-prb07} questions about its stability and about
possible pathways from the ${\alpha}$ to the ${\beta}$ phase have not been
fully explored.

In the present work, we have carried out first-principles calculations
for candidate 
${\alpha}$ and ${\beta}$ cristobalite structures in the framework of
density-functional theory (DFT) in order to check the stability of both phases
and to explore the energy landscape connecting them.  We have also
calculated phonon frequencies and infrared activities for both
${\alpha}$ and ${\beta}$ phases, and explored how the phonon modes in the
different phases are related to each other and to those of the
high-symmetry cubic phase. 
Our calculations are effectively zero-temperature ones, and thus
cannot properly treat the temperature-induced $\alpha$--$\beta$
cristobalite phase transition.  Nevertheless we hope that the
information obtained from these calculations can eventually be built into a
realistic statistical-mechanical theory that correctly takes the RUM
fluctuations into account in its description of the $\alpha$ and
$\beta$ phases at experimentally relevant temperatures.

The paper is organized as follows.  In Sec.~\ref{sec:prelim} we
give a brief review of ${\alpha}$ and ${\beta}$ cristobalite structures
and describe the methods used in the calculations. Then, in
Sec.~\ref{sec:results}, we present the results
of our calculations of
structural and lattice vibrational properties of the two phases and of
the energy landscape connecting them. We discuss those results in
Sec.~\ref{sec:discussion}. Finally, we summarize the work in Section
\ref{sec:summary}.

\section{Preliminaries}
\label{sec:prelim}

\subsection{\label{sec:introStruc}Cristobalite structures}

In order to describe the structures of the SiO$_2$ ${\alpha}$ and
${\beta}$ cristobalite phases, it is easiest to start by considering the
``ideal cristobalite'' structure, which is constructed by placing Si
atoms in a diamond structure with oxygen atoms located midway between
each pair of nearest-neighbor Si atoms.  This structure has the space
group \phIdeal\ (\phIdealSch) and has two formula units per primitive
unit cell.  Each Si atom with its four surrounding O atoms forms a
tetrahedron, so the whole structure can be visualized as a network of
tetrahedra connected at their apices.

\begin{figure}
\includegraphics{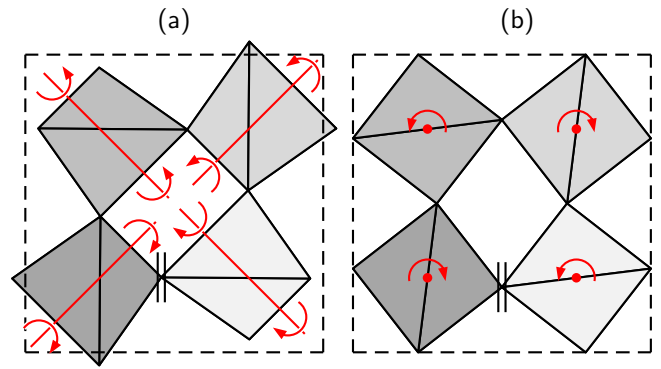}
\caption{\label{fig:rums}(Color online) Projection on $x$-$y$ plane of the
  (a) $\tilde{\alpha}$ and (b) $\tilde{\beta}$ structures,
  proposed as candidates for $\alpha$ and $\beta$ cristobalite
  phases respectively.
  Darker shading is used to represent more distant tetrahedra so
  that the spiral structure of the connected tetrahedrons
  becomes evident; double vertical lines indicate that the
  adjoining tetrahedra are actually disconnected because
  they are separated in the $z$-direction.}
\end{figure}

The generally accepted structure of $\alpha$-cristobalite is arrived
at by starting from the ideal structure and making nearly rigid
rotations of the tetrahedra about [100] and [010] axes (in the
original diamond cubic frame), leading to a tetragonal structure with
its axis along $\hat{z}$.  This is illustrated in
Fig.~\ref{fig:rums}(a), but in the conventional tetragonal frame,
related to the original cubic frame by a 45$^\circ$ rotation about
$\hat{z}$. The tetrahedral rotations are also accompanied by small
strains and tetrahedral translations needed to keep the apices
coincident, as would be expected from enforcement of the rigid-unit
constraints.  The space group of the structure is
\phAlpha\ (\phAlphaSch), and since the four rotations shown in
Fig.~\ref{fig:rums}(a) are all different, the number of formula units
per primitive unit cell is now increased to four.

As mentioned earlier, diffraction experiments on the $\beta$-cristobalite
phase tend to give inconclusive results because of spatial and dynamical
averaging.  Nevertheless, based on the comparison of structure
factors predicted by various disorder models and the ones obtained
in their x-ray diffraction experiments, Wright and
Leadbetter\cite{leadbetter} concluded that the $\beta$-cristobalite
has local \phBeta\ (\phBetaSch) space-group symmetry.  Their proposed
structure can also be
constructed from the ideal structure, but this time by rotating all
the tetrahedra around the $\hat{z}$ axis, yielding the structure shown
in \ref{fig:rums}(b). 
The number of formula units per primitive unit cell remains at two as in
the ideal structure (although it can alternatively be described, as in
Fig.~\ref{fig:rums}(b), by a doubled conventional cell containing four
formula units).  Again, the structure is highly consistent with the
rigid-unit constraints.

Because we do not want to presuppose an identification of a
particular experimentally observed {\it phase} with a particular
{\it crystal structure},
we henceforth adopt a notation in
which the phases
are identified by labels ``$\alpha$''
and ``$\beta$'' without tildes, whereas the putative crystal
structures shown in \ref{fig:rums}(a) and (b) will be referred
to as ``$\tilde{\alpha}$'' and ``$\tilde{\beta}$'' structures,
respectively.  Our working hypothesis is that the $\alpha$
and $\beta$ phases have microscopic crystal structures of type
$\tilde{\alpha}$ and $\tilde{\beta}$ respectively, but we adhere to
a distinction in the notation in order to discriminate clearly between
the specified structures used in our calculations and the hypothetical
identification of these with experimental phases.

\subsection{Computational methods}
\label{sec:methods}

The calculations were carried out using the ABINIT
implementation\cite{abinit} of density-functional theory with
Perdew-Burke-Ernzerhof\cite{ggapbe} version of the generalized
gradient approximation (GGA) for electron exchange and correlation.
Since it is the smallest unit cell that contains both $\tilde{\alpha}$
and $\tilde{\beta}$ structures, all calculations were performed on the
four-formula-unit computational
cell shown in Fig.~\ref{fig:rums}, even though the primitive cell is
smaller in the $\tilde{\beta}$ structure.
The Brillouin zone was sampled
by a $4\times4\times4$ Monkhorst-Pack grid.\cite{monkhorst}
Structural properties were computed using projector
augmented-wave\cite{paw} potentials converted from ultrasoft
pseudopotentials\cite{uspp} with a plane-wave cutoff of 22\,Ha unless
otherwise specified,
while phonon frequencies, eigenvectors, and Born charges were
computed\cite{resp} using norm-conserving Trouiller-Martins
pseudopotentials\cite{tmpseudo} at an energy cutoff of 50\,Ha
(after repeating the structural relaxation using these
potentials).
The acoustic sum rule was imposed on the force constants, and charge
neutrality was imposed on the Born charges.  Throughout the paper, the
symmetry analysis associated with crystal space groups has been
carried out using the Bilbao package.\cite{bilbao1, bilbao2}

\section{Results}
\label{sec:results}

\subsection{Structural properties of \texorpdfstring{$\bm\tilde{\bm\alpha}$}{alpha} and
  \texorpdfstring{$\bm\tilde{\bm\beta}$}{beta} structures}
\label{sec:structures}

We started our calculations by considering the ideal cubic structure
and relaxing its volume, obtaining $\ac$=7.444\,\AA\ for the lattice
constant of its eight-formula-unit cubic cell.  Then, working in the
frame of the four-formula-unit tetragonal cell, we found the
phonon frequencies at the $\Gamma$ point of its Brillouin zone,
corresponding to
phonons at the $\Gamma$ point and one X point [namely
$(2\pi/\ac)(001)$ or equivalently $(2\pi/\ac)(110)$ in the cubic
frame] of the primitive two-formula-unit fcc cell.  For the ``ideal
structure'' of space group \phIdeal, the symmetry decomposition of
these phonons into irreducible representations is
\begin{align}
  \Gamma\left(\textrm{ideal}\right) &= 1\textrm{A}_{2\textrm{u}}
  \oplus 1\textrm{E}_{\textrm{u}} \oplus 2\textrm{T}_{1\textrm{u}}
  \oplus 1\textrm{T}_{2\textrm{u}} \oplus 1\textrm{T}_{2\textrm{g}},
  \label{eq:irrepsGMIdeal} \\
  \textrm{X}\left(\textrm{ideal}\right) &= 3\textrm{X}_1 \oplus
  1\textrm{X}_2 \oplus 2\textrm{X}_3 \oplus 3\textrm{X}_4.
  \label{eq:irrepsXIdeal}
\end{align}
(The translational $\textrm{T}_{1\textrm{u}}$ mode has been omitted.)
The $\textrm{E}_{\textrm{u}}$ mode and all X modes are doubly degenerate;
the $\textrm{T}_{1\textrm{u}}$, $\textrm{T}_{2\textrm{u}}$ and
$\textrm{T}_{2\textrm{g}}$ modes are triply degenerate; and
$\textrm{A}_{2\textrm{u}}$ is non-degenerate.

\begin{figure}[t]
  \includegraphics{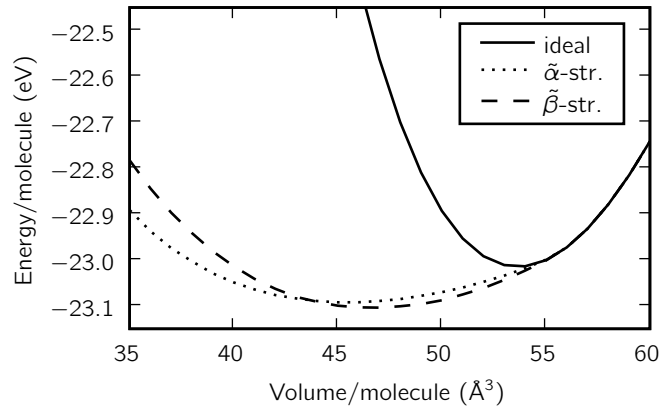}
  \caption{\label{fig:enVol} Ground state energy per formula
   unit (eV) vs.\ volume per formula unit (\AA$^3$) for $\tilde{\alpha}$,
    $\tilde{\beta}$, and cubic cristobalite structures of SiO$_2$.}
\end{figure}

We found that the triply-degenerate $\textrm{T}_{2\textrm{u}}$ mode at
$\Gamma$ is unstable with an imaginary frequency of $i 83$\,cm$^{-1}$.
All other optical phonons have real frequencies, the lowest being at
250\,cm$^{-1}$.  Furthermore, one of the doubly-degenerate
($\textrm{X}_4$) modes is unstable with a frequency of $i
53$\,cm$^{-1}$.  We thus conclude that the ideal cristobalite
structure is unstable with respect to these distortions.

Next we imposed distortions corresponding to these unstable modes and
did a full relaxation of the structure subject to the symmetry
constraints of the resulting space group. The unstable
($i53$\,cm$^{-1}$) mode at X leads to the space group \phAlpha\ 
(or \phAlphaEnan) which corresponds to $\tilde{\alpha}$-cristobalite,
while the ($i83$\,cm$^{-1}$) mode at $\Gamma$ takes us to the space
group \phBeta\ of $\tilde{\beta}$-cristobalite.  The energy of the relaxed
ground state as a function of volume per formula unit is shown for
both cases in Fig.~\ref{fig:enVol}, with the energy of the cubic phase
also shown for reference.  The corresponding structural parameters
at the energy minimum are given in Table \ref{tab:parAlphaBeta}.

\begin{table}[b]
  \caption{\label{tab:parAlphaBeta} Lattice constants (in \AA) and Wyckoff
    structural parameters for $\tilde{\alpha}$ (\phAlpha) and $\tilde{\beta}$ (\phBeta)
    cristobalite SiO$_2$.}
\begin{ruledtabular}
\begin{tabular}{lccc}
  & Present & Previous & \\
  & theory  & theory\footnotemark[1] & 
  Expt.\footnotemark[2] \\
  \hline
  \multicolumn{4}{l}{$\tilde{\alpha}$-cristobalite} \\
  \quad $a$ & 5.0730 & 5.1190 & 4.9570 \\
  \quad $c$ & 7.0852 & 7.1683 & 6.8903 \\
  \quad Si($u$)  & 0.3001 & 0.2869 & 0.3047 \\
  \quad O($x$)   & 0.2384 & 0.2439 & 0.2381 \\
  \quad O($y$)   & 0.1081 & 0.0777 & 0.1109 \\
  \quad O($z$)   & 0.1819 & 0.1657 & 0.1826 \\
  \multicolumn{4}{l}{$\tilde{\beta}$-cristobalite} \\
  \quad $a$ & 7.1050 & 7.226 & \;7.131\footnotemark[3] \\
  \quad $c$ & 7.4061 & 7.331 & \;7.131\footnotemark[3] \\
  \quad O($x$)   & 0.1051 & 0.0896 & 0.079 \\
\end{tabular}
\end{ruledtabular} 
\footnotetext[1]{Ref.~[\onlinecite{demuth}].}
\footnotetext[2]{Refs.~[\onlinecite{leadbetter}] and [\onlinecite{pluth}].}
\footnotetext[3]{Experiment sees average cubic structure.}
\end{table}

From Fig.~\ref{fig:enVol} it is clear that the $\tilde{\alpha}$ and $\tilde{\beta}$
structures indeed have lower energies than the ideal cristobalite when
the volume becomes smaller than some critical volume $V_0\sim
55\textrm{~\AA$^3$}$.  (Above this volume, the imposed distortions
disappear during relaxation and the structure returns to the ideal
one.)  We find that both the $\tilde{\alpha}$ and $\tilde{\beta}$ structures
have a quite
similar dependence of energy on volume.  According to our calculation,
the relaxed $\tilde{\beta}$ structure has a slightly lower
energy than that of the $\tilde{\alpha}$ structure (12~meV per
formula unit). This appears to be in conflict
with the experimental situation, since the $\alpha$ phase is
experimentally more stable at lower temperatures.  However, when we
repeated our calculations using a local-density approximation (LDA)
exchange-correlation functional,\cite{PZCA} the $\tilde{\beta}$ structure was
found to be lower by 1~meV per formula unit.  We thus conclude
that the small energy difference between the two nearly-degenerate
structures is a quantity that is too delicate to be reliably obtained by
our DFT calculations.  A similar discrepancy between the results from
LDA and GGA functionals was found in Ref.~[\onlinecite{demuth}].

We also analyzed what happens to the bond lengths and angles in the
$\tilde{\alpha}$ and $\tilde{\beta}$ structures as a function of volume.  The
details of this analysis are deferred to App.~\ref{app:rum}.
Briefly, for $V<V_0$, the O--Si--O bond angles and Si--O bond lengths
inside the tetrahedra are found to remain almost constant, while
the Si--O--Si bond angles change by $\sim$35$^{\circ}$.  The details
are shown to be very close to the predictions of a picture of
tilting of perfectly rigid tetrahedra.  The fact that the
three phases are indistinguishable for $V>V_0$ is also easily
explained, since the tilts of rigid tetrahedra can
only decrease the volume of the ideal structure.  Thus, for $V>V_0$
the tetrahedra cannot remain rigid and the Si--O bond length must
increase, and only when $V$ becomes smaller than some volume $V_0$ will
one of the RUMs condense in the structure in order to maintain the
preferred Si--O bond lengths.

\subsection{Phonons}
\label{sec:phonons}

\subsubsection{Phonons at \texorpdfstring{$\Gamma$ }{gamma}in \texorpdfstring{$\tilde{\alpha}$}{alpha} cristobalite}
\label{sec:phonAlpha}

\begin{table}[t]
  \caption{\label{tab:phonAlphaIrIrRa} Infrared-active phonon modes at
    $\Gamma$ in $\tilde{\alpha}$-cristobalite (\phAlpha). (E modes are also
    Raman-active.)  For A$_2$ modes,
    $\omega_{\textrm{LO}}$ refers to a phonon with
    $\hat{q}=\hat{z}$, while for E modes
    $\omega_{\textrm{LO}}$ refers to $\hat{q}$ lying in x-y plane.}
\begin{ruledtabular}
\begin{tabular}{l,c,}
   Irrep & 
   \mc{$\omega_{\textrm{\tiny TO}}$ (cm$^{-1}$)} & 
   \mc{$\tilde{\textrm{Z}}_{\lambda}$} & 
   \mc{$\omega_{\textrm{\tiny LO}}$ (cm$^{-1}$)}\\
   \hline
   E     & 127. & 0.05 & 128. \\
   E     & 259. & 0.04 & 260. \\
   A$_2$ & 285. & 0.20 & 293. \\
   E     & 357. & 0.18 & 360. \\
   E     & 440. & 0.74 & 507. \\
   A$_2$ & 462. & 0.67 & 515. \\
   E     & 584. & 0.23 & 591. \\
   A$_2$ & 751. & 0.52 & 764. \\
   E     & 752. & 0.02 & 752. \\
   A$_2$ & 1050.& 1.52 & 1201.\\
   E     & 1170.& 0.17 & 1165.\\
   E     & 1048.& 1.55 & 1208.\\
\end{tabular}
\end{ruledtabular}
\end{table}

We next repeated the calculation of the phonon frequencies for the
fully relaxed $\tilde{\alpha}$ and $\tilde{\beta}$ cristobalite structures.  We did
this in order to compare with experimental measurements and also to
check the stability of the structures and to investigate,
at least at harmonic order, the nature of the energy landscape
around these structures.  An analysis extending beyond the
harmonic approximation will be presented in Sec.~\ref{sec:saddle}.

The decomposition of the optical $\Gamma$ phonons into irreducible
representations for the $\tilde{\alpha}$ structure in space group \phAlpha\ is
\begin{equation}
  \Gamma\left(\tilde{\alpha}\right) = 4 \textrm{A}_1 \oplus 
  4 \textrm{A}_2 \oplus 5 \textrm{B}_1 \oplus 4 \textrm{B}_2 
  \oplus 8 \textrm{E}. \label{eq:irrepsGMAlpha}
\end{equation}
(The translational A$_2$ and E zero modes have been omitted.)
Only the E modes are doubly degenerate; all others are
non-degenerate.

Tables \ref{tab:phonAlphaIrIrRa} and \ref{tab:phonAlphaRa} present the
phonon frequencies at the $\Gamma$ point for the fully relaxed
$\tilde{\alpha}$-cristobalite structure.  All phonon frequencies are positive,
although some appear to be rather low in frequency.  For the
infrared (IR) active modes shown in Table \ref{tab:phonAlphaIrIrRa},
the transverse mode frequencies were computed initially, and their
mode dynamical charges were also computed using
\begin{equation}
  \tilde{\textrm{Z}}_{\lambda,\alpha}^*  =
  \displaystyle\sum_{i\beta} \frac{1}{\sqrt{M_i}} 
  \xi_{i,\lambda\beta} Z_{i,\alpha\beta}^*
  \label{eq:bornCharge}
\end{equation}
where $\xi_{i,\lambda\beta}$ is an eigenvector of the dynamical matrix,
$Z_{i,\alpha\beta}^*$ is the Born atomic charge tensor, and $M_i$ is the
mass of the $i$-th atom in amu.  The norms of the mode-charge vectors
$\tilde{\textrm{Z}}_{\lambda}^* =
  [\sum_{\alpha} ( \tilde{Z}_{\lambda,\alpha}^* )^2 ]^{1/2}$
are also given in the Table.  The longitudinal dynamical matrix
was then constructed and diagonalized using standard methods,\cite{resp}
and the resulting LO mode frequencies are presented in the last
column of Table \ref{tab:phonAlphaIrIrRa}.  It can be seen that there
are wide variations in the mode dynamical charges, and consequently,
large variations in the LO--TO splittings.

\begin{table}[!h]
  \caption{\label{tab:phonAlphaRa} Raman-only phonon modes at $\Gamma$
    in $\tilde{\alpha}$-cristobalite (\phAlpha).}
\begin{ruledtabular}
 \begin{tabular}{l,l,l,}
   Irrep &  \mc{$\omega$ (cm$^{-1}$)} & Irrep &  \mc{$\omega$
     (cm$^{-1}$)} & Irrep &  \mc{$\omega$ (cm$^{-1}$)} \\
   \hline
   B$_1$ & 29.  & B$_1$ & 358. & A$_1$ & 1046. \\ 
   B$_1$ & 103. & A$_1$ & 378. & B$_1$ & 1049. \\
   A$_1$ & 197. & B$_2$ & 410. & B$_2$ & 1109. \\
   B$_2$ & 275. & B$_1$ & 745. &       &       \\
   A$_1$ & 350. & B$_2$ & 750. &       &       \\ 
 \end{tabular}
\end{ruledtabular}
\end{table}

\subsubsection{Phonons at \texorpdfstring{$\Gamma$}{gamma} and {\rm M} in \texorpdfstring{$\tilde{\beta}$}{beta}-cristobalite}
\label{sec:phonBeta}

\begin{table}[t]
  \caption{\label{tab:phonBetaIrRa} $\Gamma$ phonons in
    $\tilde{\beta}$-cristobalite (\phBeta). In italics we show for each
    phonon in $\tilde{\beta}$-cristobalite a closest phonon in 
    $\tilde{\alpha}$-cristobalite (\phAlpha).}
  \begin{ruledtabular}
    \begin{tabular}{l,ll,}
      \multicolumn{3}{c}{Phonon in $\tilde{\beta}$ structure} & 
      \multicolumn{2}{c}{Closest in $\tilde{\alpha}$ structure}\\
      \hline
      \mc{Irrep} & \mc{$\omega$ (cm$^{-1}$)} &
      \mc{$\tilde{\textrm{Z}}_{\lambda}$} &
      \mc{Irrep} & \mc{$\omega$ (cm$^{-1}$)} \\
      \hline
      \multicolumn{5}{l}{Infrared and Raman}\\
      E     & 126.  & 0.06 & {\it E} & {\it 127}.\\
      B$_2$ & 425.  & 0.79 & {\it A$_\textit{2}$} & {\it 462}.\\
      E     & 444.  & 0.72 & {\it E } & {\it 440}.\\
      E     & 748.  & 0.51 & {\it E } & {\it 752}.\\
      B$_2$ & 1038. & 1.56 & {\it A$_{\textit{2}}$} & {\it 1050}.\\
      E     & 1047. & 1.52 & {\it E } & {\it 1048}.\\
      \multicolumn{5}{l}{Raman only}\\
      A$_1$ & 289.  & & {\it B$_\textit{1}$} & {\it 29}.\\
      B$_1$ & 406.  & & {\it A$_\textit{1}$} & {\it 350}.\\
      B$_1$ & 737.  & & {\it B$_\textit{1}$} & {\it 745}.\\
      \multicolumn{5}{l}{Inactive}\\
      A$_2$ & 357.  & & {\it B$_\textit{2}$} & {\it 410}. \\
      A$_2$ & 1097. & & {\it B$_\textit{2}$} & {\it 1109}.\\
 \end{tabular}
\end{ruledtabular}
\end{table}

\begin{table}[b]
  \caption{\label{tab:phonBetaM} M phonons in
    $\tilde{\beta}$-cristobalite (\phBeta). In italics we show for each
    phonon in $\tilde{\beta}$-cristobalite a closest phonons in 
    $\tilde{\alpha}$-cristobalite (\phAlpha).}
  \begin{ruledtabular}
    \begin{tabular}{l,l,l,}
      \multicolumn{2}{c}{Phonon in $\tilde{\beta}$ structure} & 
      \multicolumn{4}{c}{Closest in $\tilde{\alpha}$ structure}\\
      \hline
      \mc{Irrep} & \mc{$\omega$ (cm$^{-1}$)}
    & \mc{Irrep} & \mc{$\omega$ (cm$^{-1}$)}
    & \mc{Irrep} & \mc{$\omega$ (cm$^{-1}$)} \\
      \hline
      M$_3$M$_4$ & 35. &
      {\it B$_\textit{1}$} & {\it 103}.&
      {\it A$_\textit{1}$} & {\it 197}.\\
      M$_5$ & 281. & 
      {\it E } & {\it 259}.&
      {\it E } & {\it 357}.\\
      M$_1$M$_2$ & 316. &
      {\it B$_\textit{2}$} & {\it 275}.&
      {\it A$_\textit{2}$} & {\it 285}.\\
      M$_3$M$_4$ & 336. &
      {\it B$_\textit{1}$} & {\it 358}. &
      {\it A$_\textit{1}$} & {\it 378}. \\
      M$_5$ & 372. &
      {\it E } & {\it 259}.&
      {\it E } & {\it 357}.\\
      M$_5$ & 586. &
      {\it E } & {\it 584}. & & \\
      M$_1$M$_2$ & 780. &
      {\it B$_\textit{2}$} & {\it 750}. &
      {\it A$_\textit{2}$} & {\it 751}.\\
      M$_3$M$_4$ & 1045. &
      {\it A$_\textit{1}$} & {\it 1046}. &
      {\it B$_\textit{1}$} & {\it 1049}.\\
      M$_5$ & 1162. &
      {\it E } & {\it 1170}. & & \\
    \end{tabular}
  \end{ruledtabular}
\end{table}

Similar calculations of phonon frequencies were also carried out for the
$\tilde{\beta}$ structure proposed by Wright and Leadbetter\cite{leadbetter}
for $\beta$-cristobalite.
Since the primitive cell of the $\tilde{\beta}$ structure contains
only two formula units while the $\tilde{\alpha}$ structure contains four, it
should be kept in mind that the $\Gamma$ point of the $\tilde{\alpha}$
structure
maps not only into the $\Gamma$ point of the $\tilde{\beta}$ structure, but also
into a second point that would be denoted as ${\rm X} = (2\pi/\ac)(110)$
in the original fcc frame, or $(2\pi/a)(100)$ (where $a\simeq
\ac/\sqrt{2}$) in the rotated frame of Fig.~\ref{fig:rums}; we shall
refer to this as the M point in accordance with the conventional
labeling of the body-centered-tetragonal (bct) primitive cell in the latter frame.
The decompositions of the $\Gamma$ and M phonons into irreducible
representations for the $\tilde{\beta}$ structure in space group \phBeta\
are
\begin{align}
  \Gamma\left(\tilde{\beta} \right) 
  &= 1 \textrm{A}_1 \oplus 
  2 \textrm{A}_2 \oplus 2 \textrm{B}_1 \oplus 2 \textrm{B}_2 
  \oplus 4 \textrm{E}, \label{eq:irrepsGMBeta}\\
  \textrm{M}\left(\tilde{\beta} \right) 
  &= 2 \textrm{M}_1\textrm{M}_2
  \oplus 3 \textrm{M}_3\textrm{M}_4 \oplus 4 \textrm{M}_5.
\label{eq:irrepsMBeta}
\end{align}
(Translational B$_2$ and E zero modes have been omitted.)
All M-point modes and $\Gamma$-point E modes are doubly
degenerate, while other modes are non-degenerate.

In Tables \ref{tab:phonBetaIrRa} and \ref{tab:phonBetaM} we present
the our results for the $\Gamma$-point and M-point phonon modes,
respectively, of $\tilde{\beta}$-cristobalite.  The frequencies given
for the IR-active modes at $\Gamma$ are the transverse ones only.
The tables also show the correspondences between the
$\Gamma$ modes in the $\tilde{\alpha}$-cristobalite structure and the $\Gamma$
and M modes in the $\tilde{\beta}$-cristobalite structure, as determined
by comparing phonon eigenvectors.

\subsubsection{\label{sec:phRelation} 
Relation to unstable phonons in the cubic phase}

\begin{table}[t]
  \caption{\label{tab:phonRelat} Relations between unstable phonons in
    ``ideal structure'' and phonons in $\tilde{\alpha}$ and $\tilde{\beta}$ structures.}
  \begin{ruledtabular}
    \begin{tabular}{l,,l,l}
      & \multicolumn{1}{l}{Ideal} & \multicolumn{2}{c}{$\tilde{\alpha}$-cristobalite} & 
      \multicolumn{2}{c}{$\tilde{\beta}$-cristobalite} \\
      & \multicolumn{1}{l}{cm$^{-1}$} & \mc{cm$^{-1}$} & Irrep &
      \mc{cm$^{-1}$} & Irrep \\
      \hline
      $\Gamma$ & i83. & 29.  & B$_1$ & 289. & A$_1$ \\
      $\Gamma$ & i83. & 127. & E     & 126. & E \\
      $\Gamma$ & i83. & 127. & E     & 126. & E \\
      X        & i53. & 197. & A$_1$ & 35.  & M$_3$M$_4$ \\
      X        & i53. & 103. & B$_1$ & 35.  & M$_3$M$_4$ \\
    \end{tabular}
  \end{ruledtabular}
\end{table}

The triply degenerate $\Gamma$-point mode of the cubic structure
having imaginary frequency $i83\textrm{\,cm$^{-1}$}$, which condensed
to form the $\tilde{\beta}$ structure, now has positive frequencies of
289\,cm$^{-1}$ for the non-degenerate A$_1$ mode and 126\,cm$^{-1}$
for the E doublet in the $\tilde{\beta}$ structure.  This same triplet
corresponds, in the $\tilde{\alpha}$ structure, to the lowest-frequency phonon of
frequency 29\,cm$^{-1}$, which has symmetry B$_1$, and to an E doublet
at 127\,cm$^{-1}$ having almost the same frequency as in the $\tilde{\beta}$
structure. The doubly-degenerate unstable mode of the cubic structure at
$i53\textrm{\,cm$^{-1}$}$, which condensed to form the $\tilde{\alpha}$ structure,
now appears in the $\tilde{\alpha}$ structure at frequencies 197\,cm$^{-1}$ and
103\,cm$^{-1}$ with symmetries A$_1$ and B$_1$, respectively.  In the
$\tilde{\beta}$ structure, on the other hand, the same doublet appears as the
lowest-frequency phonon in that structure, namely the doublet at
35\,cm$^{-1}$ with symmetry M$_3$M$_4$.  These relations between the
unstable phonons in the ``ideal structure'' and the phonons in
$\tilde{\alpha}$ and $\tilde{\beta}$ structures are
shown in Table~\ref{tab:phonRelat}.


\subsubsection{\label{sec:phExp} Comparison with experiment for
\texorpdfstring{$\alpha$}{alpha}-cristobalite}

\begin{table*}
  \caption{\label{tab:exp} 
    Left: Computed mode frequencies and irreps for $\alpha$-cristobalite,
    with direction of dynamical polarization in parentheses for
    IR-active modes.  The modes that are adiabatically connected as
    $\hat{q}$ is rotated from $\hat{x}$ to $\hat{z}$ appear on the
    same line.  All modes other than A$_2$ modes are Raman-active.
    Right: Tentative assignments to measured mode frequencies in powder
    samples.}
    \begin{threeparttable}
    \begin{ruledtabular}
    \begin{tabular*}{0.75\textwidth}{rlrlc|,,lc}
      \multicolumn{5}{c}{Present theory} &
      \multicolumn{4}{c}{Experimental data} \\
      \hline
      \multicolumn{2}{c}{$\hat{q}\parallel\hat{x}$} &
      \multicolumn{2}{c}{$\hat{q}\parallel\hat{z}$} & &
      \multicolumn{1}{c}{IR$^{\mathrm{g}}$} & \multicolumn{1}{c}{Raman$^{\mathrm{h}}$} & & \\
      \mc{cm$^{-1}$} & Irrep & \mc{cm$^{-1}$} & Irrep & Notes &
      \mc{cm$^{-1}$} & \mc{cm$^{-1}$} & Irrep$^{\mathrm{i}}$ & Notes \\
      \hline
      1208 & E\,(x) & 1201 & A$_2$(z) & {\it a,e} & 1144 & & A$_2$ & {\it c}\\
      1170 & E\,(y) & 1170 & E\,(y) & {\it d} & 1196 & 1193 & E & {\it d}\\ 
      1165 & E\,(x) & 1170 & E\,(x) & {\it d} & & & & \\ 
      1109 & B$_2$ & 1109 & B$_2$ & {\it d} & & 1188 & B$_2$ & {\it d} \\
      1050 & A$_2$(z) & 1048 & E\,(x) & {\it e} &  & & & \\
      1048 & E\,(y) & 1048 & E\,(y) & {\it } & 1100 & - & E & {\it c} \\
      1049 & B$_1$ & 1049 & B$_1$ & {\it d} & & 1086 & A$_1$ or B$_1$ & {\it d}\\
      1046 & A$_1$ & 1046 & A$_1$ & {\it d} & & 1076 & A$_1$ or B$_1$ & {\it } \\
      751 & A$_2$(z) & 764 & A$_2$(z) & {\it d} & 798 & & & {\it }\\ 
      752 & E\,(y) & 752 & E\,(y) & {\it b} & & & & {\it }\\
      752 & E\,(x) & 752 & E\,(x) & {\it b} & & & & \\
      750 & B$_2$ & 750 & B$_2$ & {\it d} & & 796 & & {\it d}\\
      745 & B$_1$ & 745 & B$_1$ & & & 785 & B$_1$ & \\
      591 & E\,(x) & 584 & E\,(x) & {\it d} & & & & \\
      584 & E\,(y) & 584 & E\,(y) & {\it d} & 625 & - & E & {\it d} \\
      507 & E\,(x) & 515 & A$_2$(z) & {\it a,e} & & & & \\
      462 & A$_2$(z) & 440 & E\,(x) & {\it e} & 495 & & A$_2$ & {\it c} \\
      440 & E\,(y) & 440 & E\,(y) & {\it } & 480 & 485? & E & {\it c} \\
      410 & B$_2$ & 410 & B$_2$ & {\it d} & & 426 &A$_1$ or B$_2$ & {\it d} \\
      378 & A$_1$ & 378 & A$_1$ & {\it d} & & & & {\it }\\
      360 & E\,(x) & 357 & E\,(x) & {\it d} & & & & \\
      357 & E\,(y) & 357 & E\,(y) & {\it d} & 380 & 380 & E & {\it d} \\
      358 & B$_1$ & 358 & B$_1$ & {\it d} & & & & {\it }\\
      350 & A$_1$ & 350 & A$_1$ & & & 368 & A$_1$ or B$_1$ & \\
      285 & A$_2$(z) & 293 & A$_2$(z) & {\it d} & 300 & & A$_2$ & {\it d} \\
      275 & B$_2$ & 275 & B$_2$ & {\it d} & & 286 & B$_2$ & {\it d} \\
      260 & E\,(x) & 259 & E\,(x) & {\it d} & & & & \\
      259 & E\,(y) & 259 & E\,(y) & {\it d} & 276 & 275 & E & {\it d} \\
      197 & A$_1$ & 197 & A$_1$ & {\it d,f} & & 233 & A$_1$ & {\it d} \\
      128 & E\,(x) & 127 & E\,(x) & {\it f} & & & & \\
      127 & E\,(y) & 127 & E\,(y) & {\it f} & 147 & - & E & \\
      103 & B$_1$ & 103 & B$_1$ & {\it d,f} & & 121 & B$_1$ & {\it d} \\
      29 & B$_1$ & 29 & B$_1$ & {\it f} & & 50 & B$_1$ & \\
 \end{tabular*}
\begin{tablenotes}[para]
 \pitem[a] Not pure LO at all $\hat{q}$.
 \pitem[b] LO-TO splitting is negligible. 
 \pitem[c] Part of structured peak.
 \pitem[d] Inactive in $\beta$ phase.
 \pitem[e] In $\tilde{\beta}$ structure the A$_2$ component also becomes
     Raman active.
 \pitem[f] Corresponds to RUM mode in ideal cristobalite.
 \pitem[g] Observation from [\onlinecite{swainson}].
 \pitem[h] Observation of 50\,cm$^{-1}$ mode is from
   [\onlinecite{sigaev-jns99}], all others from [\onlinecite{bates}].
 \pitem[i] Empirical assignments from [\onlinecite{swainson}].
 \end{tablenotes}
 \end{ruledtabular}
 \end{threeparttable}
\end{table*}

In view of the recent questions that have been posed about the
identity of the $\alpha$-cristobalite phase,\cite{scott} we have
carried out a more detailed analysis of the phonons in the $\tilde{\alpha}$
structure.  In
particular, we have calculated the LO frequencies of the
$\Gamma$-point phonons in $\tilde{\alpha}$-cristobalite as a function of the
angle at which the limit $\hat{q} \rightarrow 0$ is taken. It turns
out that the labels A$_2$ and E are not well-defined at arbitrary
$\hat{q}$ because of mixing between modes of these symmetries.
Moreover, it can happen that if one starts with an E mode at $\hat{q}
\parallel \hat{z}$ and follows the branch as $\hat{q}$ is rotated, one
arrives at an A$_2$ mode when $\hat{q}$ lies in the $x$-$y$ plane, or
vice versa.  Experiments have typically been done on powder samples,
so that one should in principle average the phonon spectrum over all
possible directions for $q\rightarrow0$.  Moreover, some phonon modes
with E symmetry have a very small LO-TO splitting, so they would most
likely appear in experiment as a single line.

For all these reasons, a direct comparison of experimental data with
our results as presented in Tables \ref{tab:phonAlphaIrIrRa} and
\ref{tab:phonAlphaRa} is
problematic. Nevertheless, we attempt such a comparison in
Table~\ref{tab:exp}. Despite the difficulties,
the agreement with experimental data is rather good, with a
few exceptions that will be discussed shortly.  We generally
underestimate the experimental frequencies by $\sim$20\,cm$^{-1}$ for
lower frequency phonons and by $\sim$35\,cm$^{-1}$ for higher
frequencies, but otherwise our results reproduce the experimental
pattern of frequencies, and the irrep assignments are also consistent
with those obtained from empirical models.\cite{swainson}
Furthermore, the identification of the modes that are not expected
to be active in the ${\beta}$ phase (fifth and ninth columns of
Table~\ref{tab:exp}) because they correspond to zone boundary modes in
the ${\beta}$ phase (see Table~\ref{tab:phonBetaM}) or because they are
inactive $\Gamma$-point modes
(see Table~\ref{tab:phonBetaIrRa}) agrees well with the results
reported in Ref.~[\onlinecite{swainson}].

The first anomaly is related to the experimentally observed IR mode with
a frequency of 798\,cm$^{-1}$ in the $\alpha$ phase that does not
disappear upon transition to the $\beta$ phase.
Finnie {\it et al.}~[\onlinecite{finnie}]
explained this by suggesting that a two-phonon process in the
$\beta$ phase replaces the fundamental mode of the  $\alpha$ phase.
Our calculations identify two almost-degenerate IR modes that
are close to this frequency, an A$_2$ mode at
751\,cm$^{-1}$ and an E mode 752\,cm$^{-1}$.
In the $\tilde{\alpha}$ structure both of these modes are IR active,
but the Born charge of the E mode is only 0.02 while that of A$_2$ is
0.52, which means that the E mode in the $\tilde{\alpha}$ structure
is almost invisible. In the
$\tilde{\beta}$ structure, the A$_2$ mode disappears since it is no longer at
$\Gamma$. On the other hand, the E mode remains at the $\Gamma$
point and its Born charge is increased to 0.51. These results
suggest a possible explanation for the ``anomaly,'' namely that
there are two IR modes in the $\alpha$ phase; one of them is much
weaker than the other, but upon the transition to the $\beta$
phase, the stronger one disappears by symmetry while the weaker one
greatly increases its IR activity. 
The reason why the 752\,cm$^{-1}$ E phonon in the $\tilde{\alpha}$
structure acquires a larger Born charge upon converting to the
$\tilde{\beta}$ structure is that it gets some admixture of the
440\,cm$^{-1}$ E mode, which has a much larger Born charge (0.74)
than that of the 752\,cm$^{-1}$ mode (0.02).

The second anomaly is related to the 1076\,cm$^{-1}$ mode that is still
present upon the transition to the $\beta$ phase in the form of a
fairly broad feature
(see Fig. 1 in Ref.~[\onlinecite{bates}]), whereas it would be expected
to vanish by symmetry. Swainson {\it et al}.~[\onlinecite{swainson}]
attributed this mode to a possible higher-order process. We think that
it could also be related to the fact that the 1050\,cm$^{-1}$ A$_2$ mode
that is Raman inactive in the $\tilde{\alpha}$-phase becomes Raman active in
$\tilde{\beta}$-phase.

We also predict several phonon modes that are not seen experimentally,
such as the Raman-active modes at 127, 358, 378, 584, 752 and
1048\,cm$^{-1}$.  Since we have not computed Raman matrix elements, it
is possible that the Raman intensities are small for these modes.
We also find one weak IR-active mode at 752\,cm$^{-1}$ that is not
seen in the experiments.

A very low-frequency phonon at 50\,cm$^{-1}$ has been reported in
$\alpha$ cristobalite.\cite{volkov,sigaev-jns99,leadbetterN}
We believe this most
likely corresponds to the B$_1$ phonon that we have calculated to
appear at 29\,cm$^{-1}$, corresponding closely to the RUM mode
that takes the ideal cubic cristobalite
structure into the $\tilde{\beta}$ structure.
The same conclusion regarding the lowest-frequency B$_1$ phonon
was reached in Ref.~[\onlinecite{swainson}].
The minimal-energy path between $\alpha$ and $\beta$
phases that is related to this low-frequency phonon is discussed
in Sec.~\ref{sec:saddle}.

\subsection{\label{sec:stability} SiO\texorpdfstring{$_{\bm2}$}{2} cristobalite
  stability analysis}

As shown in Sec.~\ref{sec:phonons}, all calculated optical phonons in
$\tilde{\alpha}$ and $\tilde{\beta}$ cristobalites have $\omega^2>0$,
indicating that the relaxed structure is stable with respect to those
modes.  In view of the suggestion in Ref.~[\onlinecite{scott}] that
the $\alpha$ phase might locally have $D_2$ rather than $D_4$ point-group
symmetry, we checked carefully for instabilities leading from
the $\tilde{\alpha}$ structure to $D_2$ structures, but found none.
The possible subgroups of
\phAlpha\ (\phAlphaSch) having $D_2$ symmetry (without reduced
translational symmetry) are
\phAlphaScott\ (\phAlphaScottSch) and \phSubGr\ (\phSubGrSch),
and
the phonon distortions leading to these symmetry-lowered structures
are the ones of $B_2$ and $B_1$ symmetry respectively.
The lowest-frequency mode of $B_2$ symmetry is at 275\,cm$^{-1}$, so
there is certainly no sign of an instability there.
On the other hand, the lowest-frequency $B_1$ phonon is nearly soft at
29\,cm$^{-1}$, suggesting that the $\tilde{\alpha}$ structure is nearly
unstable to a spontaneous transformation into the \phSubGr\ structure.
To check this possibility more carefully, we started
from the relaxed $\tilde{\alpha}$ structure and followed
the distortion corresponding to the 29\,cm$^{-1}$ $B_1$ phonon,
and confirmed that the energy increases monotonically (no
double-well structure).  Moreover, starting from one of these
structures having a small amount of the 29\,cm$^{-1}$ mode frozen in,
a subsequent relaxation inside the resulting space group \phSubGr\
lead to a recovery of the starting
space group \phAlpha.  We thus conclude, at least within our
zero-temperature first-principles calculations, that the
$\tilde{\alpha}$ structure is locally stable, i.e., does not
spontaneously lower its symmetry to $D_2$.

Nevertheless, the presence of several modes of quite low frequency in
the $\tilde{\alpha}$-cristobalite structure may be suggestive
of low-energy pathways leading from the $\tilde{\alpha}$ to the $\tilde{\beta}$
structure or between domains of the $\tilde{\alpha}$ structure.  For example,
we have shown above that the lowest-frequency 29\,cm$^{-1}$ mode in the
$\tilde{\alpha}$-cristobalite structure corresponds to a phonon of the
``ideal structure'' that leads to the $\tilde{\beta}$ structure, and vice versa.
This might suggest that there is a relatively low energy barrier in
the configuration space that connects one structure to the other.
Other phonons from the unstable triplet and doublet in the
``ideal structure'' have frequencies that are higher, but still
low enough to suggest that there is a low energy barrier for
creation of the domains.  In order to clarify these issues, we shall
explore the energy landscape around the $\tilde{\alpha}$ and $\tilde{\beta}$
structures in more detail in Sec.~\ref{sec:saddle}.  First, however,
we begin with a general discussion of RUMs in the cristobalite
phases in the next subsection.

\subsubsection{\label{sec:rums} Rigid unit mode analysis}

\begin{figure*}
  \includegraphics{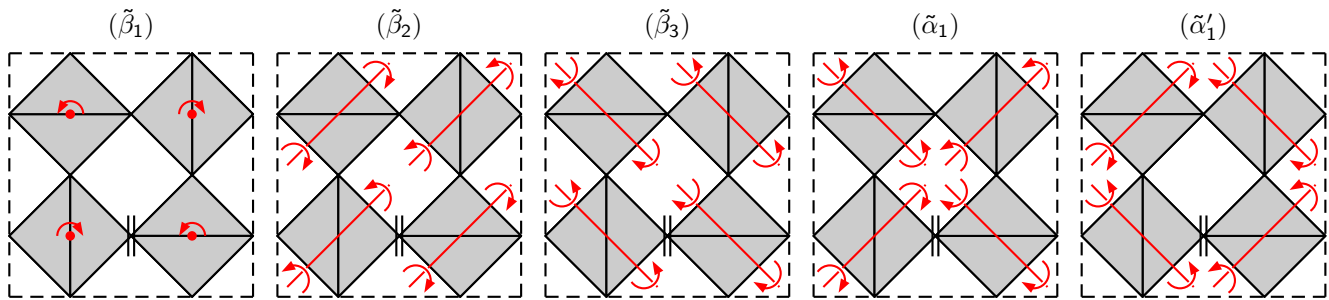}
  \caption{\label{fig:imagAll} (Color online)
  Five linearized RUMs in ideal cristobalite $Z$=4 structure.}
\end{figure*}

Here we analyze
the RUMs present in the high-symmetry cubic structure,\cite{dove-rum}
but constrained to maintain the periodicity of the four-formula-unit
($Z$=4) cell of the $\tilde{\alpha}$-cristobalite structure (i.e., containing
two unit cells of the $\tilde{\beta}$-cristobalite structure).  When these
constraints are imposed, we find that there are five linearly
independent RUM distortions as shown in Fig.~\ref{fig:imagAll}.  The
first three of these distortions, which we denote as $\tilde{\beta}_1$,
$\tilde{\beta}_2$, and $\tilde{\beta}_3$, consist of tetrahedral rotations of
alternating signs about a single Cartesian axis, and carry the system
into the \phBeta\ symmetry of the $\tilde{\beta}$ phase.  The last two, which
we denote as $\tilde{\alpha}_1$ and $\tilde{\alpha}_1'$,
consist of a pattern of
rotations around the $[110]$ and $[1\bar{1}0]$ axes
in the frame of Fig.~\ref{fig:rums}, together with small translations
of the tetrahedra needed to keep them connected at their apices, and
carry the system into the \phAlpha\ (or, for  $\tilde{\alpha}_1'$, into the
enantiomorphic
\phAlphaEnan) symmetry of the $\tilde{\alpha}$-cristobalite structure.
Not shown in Fig.~\ref{fig:rums} are RUM rotations $\tilde{\alpha}_2$
and $\tilde{\alpha}_2'$ associated with a second X point, and
$\tilde{\alpha}_3$ and $\tilde{\alpha}_3'$ associated with a third X point.

Within the context of an ideal rigid-unit geometry (in which no additional
relaxations are allowed), one can make the following mathematical
analysis.  The freezing in of the ideal $\tilde{\beta}_1$ RUM leads to a
$\tilde{\beta}$ structure oriented as in Sec.~\ref{sec:structures},
whereas the freezing in of the ideal $\tilde{\alpha}_1$ RUM leads to an
$\tilde{\alpha}$ structure as in that section.  In the ideal $\tilde{\beta}$
structure, all five of the modes shown in
Fig.~\ref{fig:imagAll} remain as true RUMs -- i.e.,
the tetrahedra can remain undistorted to first order in the mode
amplitudes.  Thus, all five modes are expected to have low
frequencies in a more realistic description.
However, modes $\tilde{\alpha}_2$, $\tilde{\alpha}_2'$, $\tilde{\alpha}_3$, and $\tilde{\alpha}_3'$ are
no longer RUMs when a finite $\tilde{\beta}_1$ RUM is present.

In the $\tilde{\alpha}_1$ structure,
only the $\tilde{\alpha}_1$, $\tilde{\alpha}_1'$, and $\tilde{\beta}_1$ distortions remain
as true RUMs.  However, the $\tilde{\beta}_2$ and $\tilde{\beta}_3$ modes at least
share the same translational symmetry, and so may be expected to have
somewhat low frequencies.  The remaining $\tilde{\alpha}_2$, $\tilde{\alpha}_2'$,
$\tilde{\alpha}_3$, and $\tilde{\alpha}_3'$ modes are incompatible both in the RUM
sense and in their translational periodicity.\cite{rumCoupl}

Not surprisingly, when we impose the translational periodicity consistent
with the five modes shown in Fig.~\ref{fig:rums},
we confirm that these five distortions correspond
quite closely to the five unstable phonon modes that we found in the ideal
structure.  The unstable $\Gamma$ modes correspond to $\tilde{\beta}_1$, $\tilde{\beta}_2$
and $\tilde{\beta}_3$, while the unstable X modes correspond to $\tilde{\alpha}_1$ and
$\tilde{\alpha}_1'$.  They also correspond closely to the low-frequency phonons
in the $\tilde{\alpha}$ and $\tilde{\beta}$ structures as discussed in
Sec.~\ref{sec:phRelation}.

Extending the mathematical analysis of the compatibility of RUMs,
it can be shown that there is an entire three-dimensional
subspace of rigid-unit structures (i.e., with the tetrahedral
rigidity condition satisfied exactly) in which finite rotations of
type ($\alpha_1$,$\alpha_1'$,$\beta_1$) are simultaneously present,
and having the space group \phSubGr\ that is induced if any two of
them are present.  In a similar way, there are two additional
three-dimensional (3D) manifolds ($\alpha_2$,$\alpha_2'$,$\beta_2$) and
($\alpha_3$,$\alpha_3'$,$\beta_3$) corresponding to different choices
of the X point and thus having different $Z$=4 supercells.  The three
subspaces meet only at a single point (the cubic phase with all angles
vanishing), and RUMs selected from different 3D manifolds are
always incompatible with each other in the sense that the perfect
tetrahedral rigidity cannot be preserved when imposing both.
This picture has important consequences for our understanding of
the possible paths connecting domains of the $\tilde{\alpha}$ and $\tilde{\beta}$
structures, as discussed below.

\subsubsection{\label{sec:saddle} Energy landscape inside 3D manifolds}

After we have explained the origin of the low-energy phonons in
the $\tilde{\alpha}$ and $\tilde{\beta}$ structures by relating them to RUM modes,
we would now like to explore the energy landscape around these
structures.  To do so, we begin by finding a configuration space
containing both structures.  Since there is no group-subgroup relation
between the $\tilde{\alpha}$ and $\tilde{\beta}$ structures, we seek a maximal common
subgroup of both structures.  In the present case, this leads to the
space group \phSubGr\ (\phSubGrSch).

In the \phSubGr\ configuration space, the $\tilde{\alpha}$ and
$\tilde{\beta}$ structures represent two special points, and we know that the
energy has local minima at these points because all computed phonon
frequencies were found to be positive there.  But then we also
expect that there must be at least one saddle point connecting
these points.  To search for this saddle point, we started from
the midpoint between the $\tilde{\alpha}$ and $\tilde{\beta}$ structures in the
12-dimensional \phSubGr\ configuration space (described by nine
internal coordinates and three cell parameters),
and identified the unit vector
$\hat{e}$ pointing between the two structures.  We then applied
a simple saddle-point search strategy in which component of the
force vector parallel to $\hat{e}$ was reversed in sign before
executing the steepest-descent update.  This algorithm can be
expected to succeed if the saddle point is not too far from the
midpoint and if the principal axis of the negative Hessian
eigenvalue at the saddle point is roughly parallel to $\hat{e}$.
In the present case, it led to a rapid convergence on the desired
saddle point.
Surprisingly, we find that the saddle point has a very low energy,
only 5~meV per formula unit above that of the $\tilde{\alpha}$ structure,
or 17~meV above that of the $\tilde{\beta}$ structure.

\begin{figure}
  \includegraphics{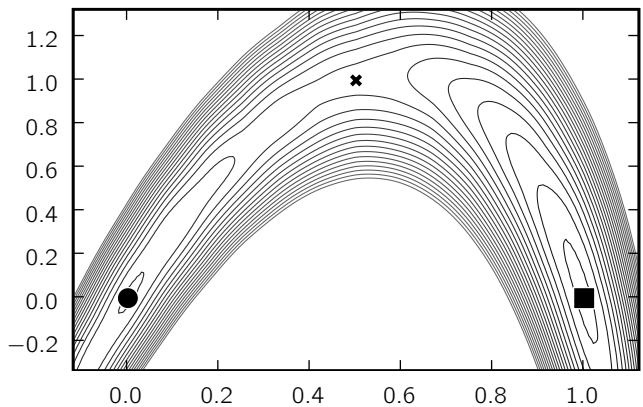}
  \caption{\label{fig:planeSaddle} Energy in plane
    defined by the $\tilde{\alpha}$ structure (filled circle),
    $\tilde{\beta}$ structure (filled square), and saddle point (cross).
    Coordinates are chosen such that these structures occur
    at $(0,0)$, $(1,0)$, and $(0.5,1)$, respectively.  The energy
    difference separating contours is 3~meV per SiO$_2$ formula unit.}
\end{figure}

The three points representing the $\tilde{\alpha}$ and $\tilde{\beta}$
structures and the saddle point determine a plane in the
12-dimensional configuration space.  To confirm that the path
running through the saddle point encounters only a single barrier,
we have plotted the structural energy
in this plane (without relaxation of other coordinates)
in Fig.~\ref{fig:planeSaddle}.
We have
somewhat arbitrarily carried out a linear transformation on the
coordinates in such a way that the $\tilde{\alpha}$ and $\tilde{\beta}$ structures
lie at $(0,0)$ and $(1,0)$ respectively, while the saddle point
lies at $(0.5,1)$, in Fig.~\ref{fig:planeSaddle}.
The results confirm the picture of a simple barrier of
5~meV encountered when going from the $\tilde{\alpha}$ to the $\tilde{\beta}$
structure.

Note that a transformation that would
lead from the $\tilde{\alpha}$ to the $\tilde{\beta}$ structure along a straight line
in configuration space would have an enormously higher barrier
of 195~meV per formula unit.  This is because the straight-line
path is a poor approximation to a RUM.  If instead we follow a
curved minimum-energy path from $\tilde{\alpha}$ through the saddle to
$\tilde{\beta}$ and compute the relaxed Si-O bond lengths and O-Si-O bond
angles along this path, we find that these remain almost constant.
This strongly suggests that this minimum-energy path may be
well approximated by some RUM-like distortion.

In the Sec.~\ref{sec:rums}, we pointed out that within the
framework of ideal rigid-unit rotations,
there is an entire three-dimensional
subspace of structures for which the tetrahedral
rigidity conditions are satisfied exactly, in which finite rotations of
all three types are present.  We label an arbitrary configuration
in this 3D manifold by ($\alpha_1$,$\alpha_1'$,$\beta_1$), where by
convention the order of operations is $\tilde{\alpha}_1$ followed by
$\tilde{\alpha}_1'$ and then $\tilde{\beta}_1$.  The space group at
a generic point in this 3D manifold is \phSubGr,
the same one we have just been discussing.  It thus seems likely
that the minimum-energy path in Fig.~\ref{fig:planeSaddle} may
correspond approximately to a path from the point ($\alpha_1$,0,0)
to the point (0,0,$\beta_1$) and lying, at least approximately, in the 
two-dimensional (2D)
subspace ($\alpha_1$,0,$\beta_1$).

To test this conjecture, we first created an ideal rigid-unit structure for
each pair of angles $(\alpha_1,\beta_1)$ on a two-dimensional mesh.
We then used our first-principles calculations to relax
each structure subject to the constraint that these
two angle variables should not change.  Technically, we did this by
carrying out the minimization of the energy in the ten-dimensional
subspace orthogonal to the two-dimensional surface for each
starting point $(\alpha_1,\beta_1)$.
We typically found that these relaxations were small,
confirming the approximate validity of the RUM picture.

\begin{figure}
  \includegraphics{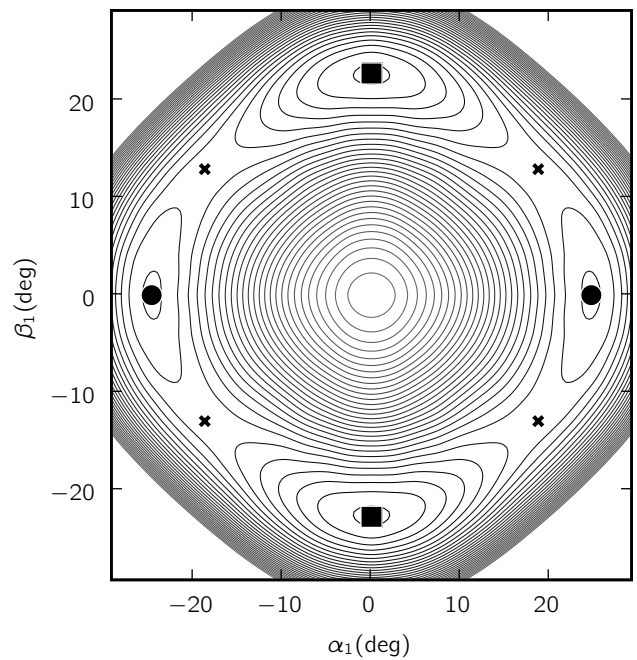}
  \caption{\label{fig:planeA1B1} Energy as a function of rotation
    angles $\phi_{\alpha_1}$ and $\phi_{\beta_1}$, corresponding to
    rotations $\tilde{\alpha}_1$ and $\tilde{\beta}_1$
    shown in Fig.~\ref{fig:imagAll}. The origin corresponds to
    ``ideal'' cristobalite. Filled squares at top and bottom denote
    $\tilde{\beta}$ minima, filled circles at left and right denote
    $\tilde{\alpha_1}$ minima, and crosses denote saddle points, as in
    Fig.~\ref{fig:planeSaddle}.  The energy
    difference separating contours is 3~meV per SiO$_2$ formula unit.}
\end{figure}

The energy surface determined in this way is plotted as a function
of rotation angles $\alpha_1$ and $\beta_1$ in Fig.~\ref{fig:planeA1B1}.
The minima corresponding to the $\tilde{\alpha}$ structure are immediately
visible near the left and right sides of the figure, while those
corresponding to the slightly lower-energy $\tilde{\beta}$ structure appear near
the top and bottom.  The minimum-energy path appears to be roughly
circular on this plot, and four equivalent saddle points are apparent
at $\alpha_1\simeq\pm19^\circ$ and $\beta_1\simeq\pm13^\circ$.
These saddle points are equivalent to the one identified in
Fig.~\ref{fig:planeSaddle}, with a barrier height of 5~meV per
formula unit relative to the $\tilde{\alpha}$ structure.  We thus confirm
the presence of a very low-energy barrier between these structures,
and identify it as approximating a certain path in the space
of rigid-unit rotations.
A video animation showing the evolution of the structure along this
path is provided in the supplementary material.\cite{EPAPS}

It is important to note that, according to the simplified model
of Eq.~(1) of Ref.~[\onlinecite{dove}],
the energy would
remain exactly zero on the entire ($\alpha_1,\beta_1$) surface of
Fig.~\ref{fig:planeA1B1} since the ideal rigid-unit structures satisfy the
rigidity conditions analytically. The RUM framework
envisages extensions to make the model more realistic; one way to
do this is by adding an energy term that depends on the relative
tilts of neighboring tetrahedra.\cite{splitSi}
We tried this by introducing a simple double-well potential model
that penalizes the departure of the Si-O-Si bond angles from a
preferred bending angle.  In this model the change of total energy
per formula unit is
\begin{equation}
 \Delta E= \frac{E_0}{N} \sum_{i} \left[-2
   \left(\frac{\pi-\phi_i}{\pi - \phi_0}\right)^2
 + \left(\frac{\pi-\phi_i}{\pi - \phi_0}\right)^4 \right]
,
\label{eq:enModel}
\end{equation}
where the sum runs over all $N$ Si-O-Si bond angles $\phi_i$ in the unit
cell. We found that we could obtain an optimal fit\cite{explan-fit}
to the results of our first-principles calculations using parameters
$E_0=83$~meV per formula unit and $\phi_0=145^{\circ}$.
The energy landscape of the fitted model looks very similar to the
results plotted in Fig.~\ref{fig:planeA1B1}. In particular, the
overall circular aspects of the energy landscape and minimal-energy
path in Fig.~\ref{fig:planeA1B1} are reproduced. However, the
model unfortunately
assigns identical energies to the $\tilde{\alpha}_1$ and $\tilde{\beta}_1$
structures, and moreover predicts a path connecting them on which
the energy remains completely flat.  This happens because, for any
given pair of angles ($\alpha_1$,$\beta_1$) on or near this path,
one can find a small $\tilde{\alpha}_1'$ such that the rigid-unit structure
($\alpha_1$,$\alpha_1'$,$\beta_1$) has all its Si-O-Si bond angles
exactly equal to $\phi_0$. Therefore our simplified model of
Eq.~(\ref{eq:enModel}), or any other model that depends solely
on the Si-O-Si angles, predicts a zero-barrier path between
$\tilde{\alpha}_1$ and $\tilde{\beta}_1$ structures.
This behavior is reminiscent of an early model of
Nieuwenkamp\cite{Nieuwenkamp} for $\beta$ cristobalite, in which
the Si-O-Si bond was assumed to rotate freely on an annulus lying
in the plane that is equidistant between Si atoms.

\begin{figure}
  \includegraphics{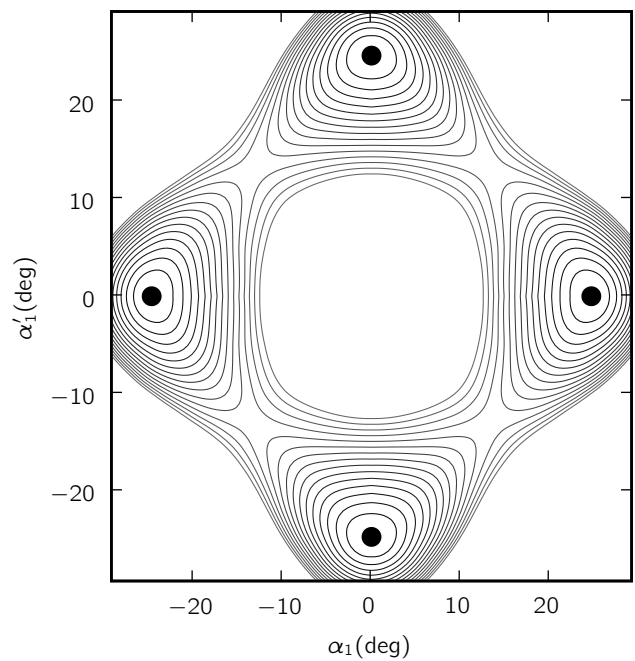}
  \caption{\label{fig:planeA1A2} Energy as a function of rotation
    angles $\phi_{\alpha_1}$ and $\phi_{\alpha_1'}$, corresponding
    to rotations $\tilde{\alpha}_1$ and $\tilde{\alpha}_1'$
    shown in Fig.~\ref{fig:imagAll}. The origin corresponds to
    ``ideal'' cristobalite.  The four minima (filled circles)
    correspond to the $\tilde{\alpha}_1'$ structure (top and
    bottom) and to the $\tilde{\alpha}_1$ structure (left and
    right).  The energy difference separating contours is 3~meV
    per SiO$_2$ formula unit.}
\end{figure}

To further test the model of Eq.~(\ref{eq:enModel}), we performed
first-principles calculations on a mesh of (unrelaxed) structures in the 2D
space ($\alpha_1$,$\alpha_1'$,$\beta_1$=0).  The resulting energy
landscape is shown in Fig.~\ref{fig:planeA1A2}.
The barrier
between $\tilde{\alpha}_1$ and $\tilde{\alpha}_1'$ structures is now about 35~meV,
substantially higher than for the path connecting $\tilde{\alpha}$
and $\tilde{\beta}$ structures.  When we use the same fitting
parameters obtained earlier, we again get very good overall
agreement; the energy landscape obtained from our model has the
same diamond-like appearance as in Fig.~\ref{fig:planeA1A2}, and
saddle points appear in very similar locations.
Moreover, the barrier of 41\,meV predicted by the model is in quite
good agreement with the first-principles value of 35\,meV.
However, in this case the picture presented by Fig.~\ref{fig:planeA1A2}
is somewhat misleading, because it turns out that the entire
minimum-energy path lying in the $\beta_1$=0 plane is unstable, and
falls to lower energy as $\tilde{\beta}_1$ is turned on.  Thus, the apparent
saddle points in Fig.~\ref{fig:planeA1A2} are actually stationary
points with {\it two} negative eigenvalues in the 3D
($\alpha_1$,$\alpha_1'$,$\beta_1$) space.  Within the model of
Eq.~(\ref{eq:enModel}), in fact, the lowest-energy path connecting
the $\tilde{\alpha}_1$ and $\tilde{\alpha}_1'$ structures is actually completely flat,
being composed of a segment connecting $\tilde{\alpha}_1$ to $\tilde{\beta}_1$ and
then another connecting $\tilde{\beta}_1$ to $\tilde{\alpha}_1'$.
This observation agrees with our first-principles calculations,
since if we start from the purported saddle-point configuration
and do a structural relaxation subject to the constraint that
$\alpha_1=\alpha_1'$, the structure is found to converge to the
$\tilde{\beta}_1$ structure as expected.

\subsubsection{\label{sec:latticeSaddle} Cell volume at minima and saddle point}

Because we have found the unit-cell parameters to be very sensitive
to details of the calculation, we increased the energy cutoff from
22\,Ha to 30\,Ha in order to obtain an accurate description of the
volume changes along the minimum-energy path.
We obtain a volume per formula unit of 45.7 and 46.7\,\AA$^3$
for the $\tilde{\alpha}$ and $\tilde{\beta}$ local minima
respectively, so that the volume is about 2.2\% larger for the
latter.  This is in qualitative agreement with experiments, which
show that the $\beta$ structure is about 5\% larger\cite{schmahl},
and implies that
applied pressure would tend to favor the $\tilde{\alpha}$ phase
and raise the $\alpha$-to-$\beta$ transition temperature.
At the saddle point, we find that the volume per formula unit
is 46.8\,\AA$^3$, which is just slightly
larger than for either of the parent-phase
structures.  This finding may be of interest for future studies of
the pressure-dependence of the phase-transition mechanism.

\subsubsection{\label{sec:dw} Domain walls}

The barriers discussed in Sec.~\ref{sec:saddle}
refer to transformation pathways in which
the crystal remains periodic and transforms homogeneously, and the
energy barriers are given per unit cell.  It would also be of
interest to consider the energies of domain walls between
various $\tilde{\alpha}$ and $\tilde{\beta}$ structures.
This is beyond the scope of the present investigation, but the
results for homogeneous transformations may give some hints as to what
could be expected.  For example, we speculate that domain walls
connecting $\tilde{\alpha}$ and $\tilde{\beta}$ structures belonging
to the same 3D rigid-unit manifold will probably have a rather low energy
per unit area, while those connecting structures belonging to
different 3D manifolds would be expected to have much higher
energies.

\section{Discussion}
\label{sec:discussion}

In this section we give a brief overview of several previously
proposed models of $\alpha$ and $\beta$ cristobalite phases, and
discuss how the results of our calculations relate to those models.

The RUM model of Ref.~[\onlinecite{dove}] describes the $\beta$ phase
as an average cubic structure that has strong dynamical fluctuations
occurring simultaneously into RUMs in all allowed regions of the
Brillouin zone. A simplified version of this picture would
be one in which the tetrahedra are assumed to be
completely free to pivot around their apices, as in
Eq.~(1) of Ref.~[\onlinecite{dove}].  In general, the simultaneous
excitation of more than one RUM will have an associated energy cost
because the tetrahedra typically cannot
remain perfectly rigid while undergoing both kinds of distortion
simultaneously.  However, as an exception, we have identified 3D
rigid-unit manifolds within which the geometrical constraints
{\it can} simultaneously be satisfied.  Within the model of
Eq.~(1) of Ref.~[\onlinecite{dove}], or the split-atom
model of Ref.~[\onlinecite{split}], the energy landscape within
this special 3D manifold would be completely flat, and one would
expect that freezing in of one RUM of type $\tilde{\alpha}_1$,
$\tilde{\alpha}_1'$ or $\tilde{\beta}_1$ would have no
consequence on the energy profile of one of these other
RUM distortions.

However, once one goes beyond the simplest versions of the model
and includes terms that depend on the Si-O-Si bond angles at the
apices, our calculations indicate that the RUM
distortions of type $\tilde{\alpha}_1$, $\tilde{\alpha}_1'$ and
$\tilde{\beta}_1$ become coupled and have a rich energy landscape.
As a step in this direction, the more sophisticated split-atom
model having an additional energy term depending on Si-O-Si bond
angles\cite{splitSi} should provide an improved description.
However, we note that even this model, or any model based solely on
Si-O-Si bond angles, still has a nonphysical behavior in that it
would necessarily predict zero-energy barriers between the
$\tilde{\alpha}_1$ and $\tilde{\beta}_1$ structures, as discussed at
the end of Sec.~\ref{sec:saddle}. Nevertheless, we believe that the
split-atom and similar models can provide important complementary
information to ours, since they are not restricted to periodic supercell
structures as ours are.

Among the models of cristobalite phase transitions is also the model
of Hatch and Ghose.\cite{hatch} They argue that the $\beta$ phase
is dynamically and spatially fluctuating between the 12 different
possible $\tilde{\alpha}$ domains having \phAlpha\ space-group symmetry.
The counting arises because there are three different X
points; each exhibits a doublet of degenerate modes leading
to enantiomorphic $\tilde{\alpha}_1$ and $\tilde{\alpha}_1'$
structures (see Fig.~\ref{fig:imagAll}); and the tetrahedra can
rotate by $\pm\phi$.  The model is based on symmetry arguments and
assumes that all of the barriers separating these 12 $\tilde{\alpha}$
structures are small.
However, our work suggests that the barriers
separating different types of $\tilde{\alpha}$ domains have
very different barriers.
Furthermore, their model does not take into account the fact that the
$\tilde{\beta}$ structure is easily accessible with a very low
barrier, suggesting that fluctuations into the $\tilde{\beta}$
structure may be more important than some of the other $\tilde{\alpha}$
structures.

Finally, O'Keeffe and Hyde\cite{okeeffe} do discuss
a path connecting $\tilde{\alpha}$ and $\tilde{\beta}$ structures,
but it is of a different type than those discussed above since
it connects $\tilde{\alpha}$ and $\tilde{\beta}$ structures
belonging to different 3D rigid-unit manifolds.  In our notation, their
path would connect $\tilde{\alpha}_1$ or $\tilde{\alpha}_1'$ to
$\tilde{\beta}_2$ or $\tilde{\beta}_3$, etc.  Such a path would
involve the simultaneous application of RUM rotations that are
incompatible with each other, and as such would be expected to have
a high energy barrier.

\begin{figure}
  \includegraphics[width=2.6in]{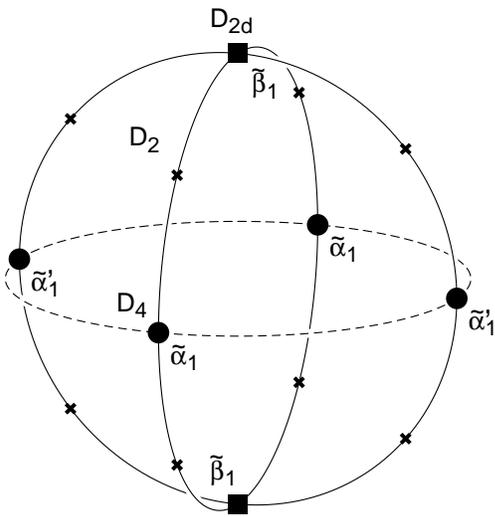}
  \caption{\label{fig:sphere} 
    Sketch of important states in one of the three-dimensional
    rigid-unit subspaces discussed in the text.  Local energy minima
    associated with $\tilde{\alpha}$ ($D_4$) and $\tilde{\beta}$
    ($D_{2d}$) structures are indicated by filled circles and
    squares respectively.  Remainder of space, including saddle
    points (crosses), has $D_2$ symmetry.}
\end{figure}

To clarify our view of the cristobalite phase transitions, we
start by emphasizing once again the existence of three distinct
3D rigid-unit manifolds, as described above at the end of Sec.~\ref{sec:rums}.
To review, one of these is described by rotation angles
($\alpha_1$,$\alpha_1'$,$\beta_1$) giving rise to structures
of space group \phSubGr\ whose translational periodicity is that
corresponding to the X point $(2\pi/\ac)(001)$ or equivalently
$(2\pi/\ac)(110)$.  This manifold contains the $\tilde{\alpha}_1$,
$\tilde{\alpha}_1'$, and $\tilde{\beta}_1$ structures, and their
partners with reversed sense of rotation, as shown schematically
in Fig.~\ref{fig:sphere}.
The second and third 3D subspaces are described by rotations
($\alpha_2$,$\alpha_2'$,$\beta_2$) and
($\alpha_3$,$\alpha_3'$,$\beta_3$), with periodicities set by
X points $(2\pi/\ac)(010)=(2\pi/\ac)(101)$
and $(2\pi/\ac)(100)=(2\pi/\ac)(011)$,
respectively.  We have found that these three subspaces are
essentially incompatible, in the sense that it is not possible to
combine rotations taken from any two of them into a combination
that preserves the rigid-unit constraints.  This occurs in part because
these three 3D rigid-unit subspaces have incompatible translational
symmetries, but also because of incompatibilities in the patterns
of rotations.

The structure of the space sketched in Fig.~\ref{fig:sphere} is intended
to reflect a three-level hierarchy of energies and energy barriers
as suggested by our analysis.
In the model of Eq.~(\ref{eq:enModel}), the energy is degenerate for
all six of the structures shown in Fig.~\ref{fig:sphere}, as well as on
the solid curves connecting them.\cite{EPAPS} According to our first-principles
results, this picture is modified so that the $\tilde{\alpha}$
and $\tilde{\beta}$ structures are local minima, with low-energy
saddle points ($\sim$5\,meV) between them (see Fig.~\ref{fig:planeA1B1}).
The low curvature of the energy surface along these curves is
reflected in the presence in Table \ref{tab:phonRelat} of a very
soft 29\,cm$^{-1}$ B$_1$ mode starting
from the $\tilde{\alpha}$ structure, and a 35\,cm$^{-1}$
M$_3$M$_4$ doublet starting from the $\tilde{\beta}$
structure.\cite{explan-dashed}  While our calculations
have the $\tilde{\alpha}$ structures at a slightly higher energy
than the $\tilde{\beta}$ ones, this is presumably reversed in the
true physical system.

The next energy scale in the hierarchy is that
associated with the direct paths between $\tilde{\alpha}$ structures
in the same 3D manifold,
indicated by the dashed lines in Fig.~\ref{fig:sphere}.  As shown in
Fig.~\ref{fig:planeA1A2}, this energy is
on the order of $\sim$35\,meV, so that the true minimum-energy
path between neighboring $\tilde{\alpha}$ structures goes instead through
(or perhaps nearly through) the $\tilde{\beta}$ structures.

Finally, the highest energies are associated with the barriers
separating any of the structures in Fig.~\ref{fig:sphere} from any
of the structures in the other two 3D subspaces. These barriers are
on the order of 80\,meV, the energy needed to pass through the
undistorted cubic phase.  While not enormously larger than the
35\,meV mentioned above, this is high enough that we do not
expect these barriers to be especially relevant for the phase
transitions in this system.

We can now speculate on the nature of the phase transition between
$\alpha$ and $\beta$ cristobalites.  We propose that
in the lower-temperature $\alpha$ phase, the system is locally
frozen onto one of the minima of type $\tilde{\alpha}$ in one
of the 3D manifolds, but with substantial fluctuations along the
low-energy paths leading to the two neighboring $\tilde{\beta}$
structures in the same manifold.  Then, in the higher-temperature
$\beta$ phase, we speculate that the system instead shifts over and
condenses locally onto one of these $\tilde{\beta}$ structures, but
with substantial fluctuations along the low-energy paths leading
to the four neighboring $\tilde{\alpha}$ structures, all in the
same 3D manifold.
The fact that there are four low-energy paths
to fluctuate along, instead of two, is consistent with the fact
that the $\beta$ phase (being the higher-temperature phase) has
higher entropy.  If the system were truly to freeze onto a single
$\tilde{\beta}$ structure, it would be globally tetragonal, with
space group \phBeta.  However, it is also possible that the system
forms on some larger scale into spatiotemporal domains composed
of $\tilde{\beta}$ structures from all three of the 3D manifolds,
giving an overall average \phIdeal\ structure in accord with the
picture espoused in Refs.~[\onlinecite{dove,dove-comm}].

Let us return for a moment to the recent work of Zhang and
Scott,\cite{scott} who argued that their Raman studies of single
crystals of $\beta$-cristobalite were inconsistent with $O_h$
symmetry.  Assuming $D_{2d}$ symmetry instead for the $\beta$
phase, these authors then noted that $D_4$ is not a subgroup
of $D_{2d}$, and thus that the existence of a group-subgroup
relation for the phase transition would rule out the assignment
of the $\alpha$ phase to the $D_4$ $\tilde{\alpha}$ structure.
On this basis, they suggested that a lower symmetry, such as $D_2$,
should be considered for $\alpha$-cristobalite.  Our view, instead,
is that a group-subgroup relation does not have to hold for
the transition, since the transition is known to be of first order,
and thus assignments of $D_{2d}$ and $D_4$ for the $\alpha$ and
$\beta$ phases respectively are not inconsistent.  As pointed out
in the introduction, while certain spectroscopic signatures of the
transition are indicative of a weakly first-order transition, the
volume change and latent heat at the transition are substantial.
The transition may perhaps be described as a reconstructive transition in the
sense of Tol\'edano and Dmitriev,\cite{toledano} although in the
present case the rearrangements of atoms can occur very gently,
because of the existence of very low-barrier paths of  $D_2$
symmetry connecting the $D_4$ ($\tilde{\alpha}$) and $D_{2d}$
($\tilde{\beta}$) structures.  The situation may be somewhat analogous
to the tetragonal--to-orthorhombic and orthorhombic-to-rhombohedral
transitions in ferroelectric perovskites such as BaTiO$_3$ and
KNbO$_3$, where the presence of low-barrier paths of monoclinic
symmetry is associated with the weakly first-order nature of the
transitions.\cite{scott-private}

Unfortunately our calculations are carried out at 0\,K with crystal
periodicity imposed.  It is therefore difficult to draw any firm
conclusions about the nature of the phase transitions between
cristobalite phases, especially if fluctuations are as important
as we think they are, and much of what we have said above must
remain speculative.  Nevertheless we hope that the results of
our calculations will be of use in developing improved models
that may allow for realistic finite-temperature modeling
of the phase transitions in this system, ultimately leading to
a resolution of the controversies that have surrounded this system
over the years.

\section{Summary}
\label{sec:summary}

Based on first-principles calculations, we
have performed a detailed analysis of the $\tilde\alpha$ (\phAlpha) and
$\tilde\beta$ (\phBeta) structures of cristobalite SiO$_2$. In particular,
we have confirmed that both structures are locally stable against all
possible distortions associated with $\Gamma$-point modes of the
four-formula-unit conventional cell.
We have calculated phonon frequencies for the $\tilde\alpha$ and
$\tilde\beta$ structures, compared these to the experimental values,
and discussed how the phonons in these two structures are related to each
other. We have also tried to resolve some experimental anomalies that were
found in spectroscopic studies of the cristobalite phases.
Finally, we have explored the energy landscape connecting the
$\tilde\alpha$ and $\tilde\beta$ structures.  We have emphasized
the existence of three distinct 3D manifolds of structures, each of
which contains both $\tilde\alpha$ and $\tilde\beta$ structures
that can be connected to each other within the manifold by paths
with a surprisingly small barrier of 5~meV per formula unit, while
paths connecting different manifolds have a much higher barrier.
While our calculations do not properly treat fluctuations, we
nevertheless have speculated on the possible consequences of our
findings for the understanding of the $\alpha$-$\beta$ phase
transition in cristobalite SiO$_2$.


\acknowledgments

This work was supported NSF Grant DMR-0549198 and ONR Grant N00014-05-1-0054.
We acknowledge useful discussions with J. F.~Scott, M. T.~Dove, K. M.~Rabe and
D. R.~Hamann.


\appendix
\section{\label{app:rum} Comparison with rigid-unit geometry}

\begin{figure}[t]
\includegraphics{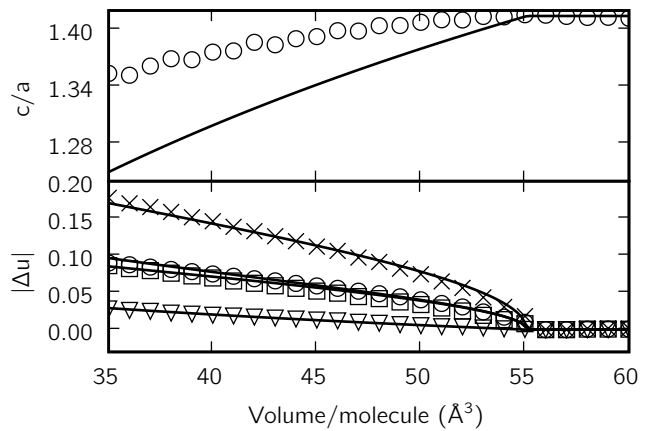}
\caption{\label{fig:fitAlpha}Structural parameters of the $\tilde{\alpha}$
  structure vs.\ volume per formula unit.  Top panel: $c/a$
  ratio.  Bottom panel: absolute values of deviations of internal
  parameters $u$(Si) (squares), $x$(O) (triangles), $y$(O)
  (crosses), and $z$(O) (circles) from ideal-cubic values.
  Symbols represent first-principles
  calculations; lines are fits to an ideal rigid-unit geometry.}
\end{figure}

In a picture in which the rigid-unit constraints are perfectly enforced,
it turns out that the structures of $\tilde{\alpha}$ and
$\tilde{\beta}$ symmetries are completely determined by a single
parameter, which can be taken to be the volume $V$ per formula unit
relative to the corresponding value $V_0$ in the ideal cubic
structure.  (That is, $V_0$ is the volume below which rigid
distortions start to appear, as explained in
Sec.~\ref{sec:structures}.)  In this Appendix, we check to see how
closely our structures, as optimized from the first-principles
calculations, match with this elementary model.

\begin{figure}[b]
\includegraphics{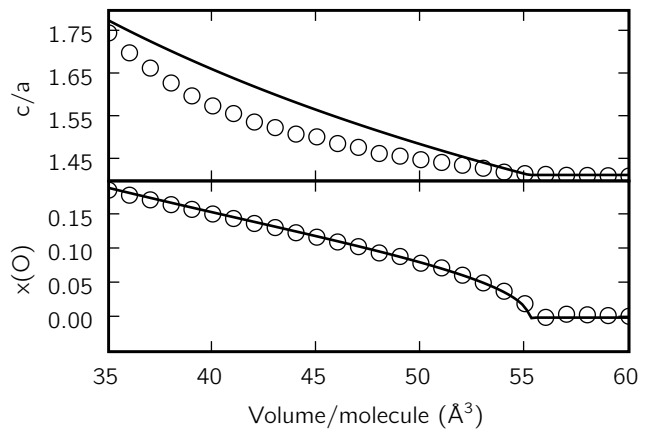}
\caption{\label{fig:fitBeta}Structural parameters of the $\tilde{\beta}$
  structure vs.\ volume per formula unit.  Top panel: $c/a$
  ratio.  Bottom panel: $x$(O).
  Symbols represent first-principles
  calculations; lines are fits to an ideal rigid-unit geometry.}
\end{figure}

The solid curves in Figs.~\ref{fig:fitAlpha} and \ref{fig:fitBeta}
show the mathematical predictions of this elementary model,
obtained by applying rotations of types $\tilde{\alpha}_1$ and
$\tilde{\beta}_1$ (see Fig.~\ref{fig:imagAll}) in such a way as to
keep the tetrahedra perfectly rigid.  (For $V>V_0$, the elementary
model cannot be satisfied, and the ideal cubic parameters are
plotted instead.)  The symbols shown in
Figs.~\ref{fig:fitAlpha} and \ref{fig:fitBeta} denote the results
of our first-principles calculations where, for each specified
value of $V$, the volume was treated as a constraint while
all other structural parameters were relaxed.  The fit was
optimized by choosing a common $V_0=55.1$~\AA$^3$
for both
$\tilde{\alpha}$ and $\tilde{\beta}$ structures.  For reference, the
first-principles equilibrium volumes are 45.7 and 46.7\,\AA$^3$ for
the $\tilde{\alpha}$ and $\tilde{\beta}$ structures, respectively.

We find that the agreement is extraordinarily good for all of the
internal parameters, but that there are some significant discrepancies
in the $c/a$ ratios.
At first sight this may seem contradictory: why are
the $c/a$ ratios off by many percent, while the Si-O
bond lengths agree within $\sim$0.05\%?  The answer is connected
with the presence of volume-preserving tetragonal distortions of
low energy cost.  In such a distortion, each tetrahedron is
stretched slightly along $c$ and compressed in $a$ (or vice
versa), and it happens that the tetrahedral angle of $\arccos(1/\sqrt{3})$
is precisely the one at which Si-O bond lengths are preserved
to first order in the distortion amplitude.  While the
O-Si-O bond angles do change at first order, this may entail a
smaller energy cost than for bond-length changes.
As expected from this analysis, we find that our first-principles
O-Si-O bond angles differ from the ideal ones by $\sim$4\%.
In short, it appears that it is energetically more important to
preserve bond lengths than bond angles, and that for geometrical
reasons this translates into an enhanced freedom for the $c/a$ ratio to
deviate from the ideal rigid-unit geometry.

\bibliography{sio2}

\begin{thebibliography}{57}
\bibitem[{exp({\natexlab{a}})}]{explan-symm}
\bibinfo{note}{Note that this is a different $D_2$ space group than the
  \phSubGr~(\phSubGrSch) one discussed elsewhere in this paper.}
\bibitem[{exp({\natexlab{b}})}]{explan-fit}
\bibinfo{note}{The fitting was done to minimize the RMS error of four
  quantities, namely the equilibrium angles and the energy difference (relative
  to the symmetric structure) for the $\alpha$ and $\beta$ structures.}
\end{thebibliography}

\end{document}